\documentclass[iop,usenatbib]{emulateapj}

\def\1{\c{c}}
\def\2{\c{C}}
\def\3{\.{I}}
\def\4{\"{a}}
\def\5{{\i}}
\def\6{$\beta$}
\def\7{\"{o}}
\def\8{\"{O}}
\def\9{\c{s}}
\def\0{\c{S}}
\def\*{\"{u}}
\def\?{\"{U}}
\def\;{\u{g}}
\def\:{\u{G}}
\usepackage{color}

\slugcomment{The Astrophysical Journal Draft Version June 11,
2014}

\shorttitle{{\it Suzaku} observation of G352.7$-$0.1}
\shortauthors{Sezer and G\"{o}k}

\begin{document}

\title{Fe-rich ejecta in the
supernova remnant G352.7$-$0.1 with {\it Suzaku}}

\author{A. Sezer*}
\affil{T\"UB\.ITAK Space Technologies Research Institute, ODTU
Campus, Ankara, 06531, Turkey} \email{* aytap.sezer@gmail.com}
\and
\author{F. G\"{o}k}
\affil{Akdeniz University, Faculty of Education, Department of Secondary Science and Mathematics Education, Antalya, 07058, Turkey\\
}
\begin{abstract}
In this work, we present results from a $\sim$201.6 ks observation
of G352.7$-$0.1 by using the X-ray Imaging Spectrometer onboard
{\it Suzaku} X-ray Observatory. The X-ray emission from the
remnant is well described by two-temperature thermal models of
non-equilibrium ionization with variable abundances with a column
density of $N_{\rm H}$ $\sim$ 3.3$\times$10$^{22}$ cm$^{-2}$. The
soft component is characterized by an electron temperature of
$kT_{\rm e}$ $\sim$ 0.6 keV, an ionization time-scale of $\tau$
$\sim$ 3.4$\times$10$^{11}$ cm$^{-3}$ s, and enhanced Si, S, Ar,
and Ca abundances. The hard component has $kT_{\rm e}$ $\sim$ 4.3
keV, $\tau$ $\sim$ 8.8$\times$10$^{9}$ cm$^{-3}$ s, and enhanced
Fe abundance. The elemental abundances of Si, S, Ar, Ca, and Fe
are found to be significantly higher than the solar values that
confirm the presence of ejecta. We detected strong Fe K-shell
emission and determined its origin to be the ejecta for the first
time. The detection of Fe ejecta with a lower ionization
time-scale favor Type Ia origin for this remnant.
\end{abstract}


\section{Introduction}
The Galactic supernova remnant (SNR) G352.7$-$0.1 was discovered
by \citet{clark1975} in the radio band at 408 and 5000 MHz
observations. The remnant has a shell structure with a radio
spectral index of $\alpha$=$-0.6$ (energy flux
S$_{\nu}\sim\nu^{\alpha}$) and an angular size of 8$\times$6
arcmin$^{2}$ \citep{green2009}. \citet{dubner1993} showed the
presence of structure of two overlapping rings by using Very Large
Array (VLA) image at 1465 MHz.

In the X-ray band, G352.7$-$0.1 was first detected by
\citet{kinugasa1998} with {\it ASCA} Galactic Plane Survey
Project. The spectra were described by a non-equilibrium
ionization (NEI) model with an electron temperature  $kT_{\rm e}$
of $\sim$ 2.0 keV, an ionization parameter $\tau$ of $\sim$
10$^{11}$ cm$^{-3}$ s, an absorption $N_{\rm H}$ of $\sim$
2.9$\times$10$^{22}$ cm$^{-2}$, and overabundances of Si and S.
Using the Sedov model and assuming a distance of 8.5 kpc (near to
the Galactic center), they estimated an explosion energy of
$\sim$2$\times$10$^{50}$ erg and an age of $\sim$2200 yr.

\citet{giacani2009} analyzed the {\it XMM-Newton} observation and
reprocessed VLA archival data at 1.4 and 4.8 GHz of G352.7$-$0.1.
They estimated a distance of 7.5$\pm $0.5 kpc by studying the
interstellar gas surrounding the SNR. Also, using {\it XMM-Newton}
data, they showed the presence of supernova (SN) ejecta in this
SNR. The spectra were well fitted with a NEI model with an
electron temperature $kT_{\rm e}$ of $\sim$1.9 keV. From spectral
fitting they estimated the electron density of the plasma to be
$n_{\rm e}$ $\sim$ 0.3 cm$^{-3}$, the age of the remnant to be
4700 yr, the mass of the X-ray emitting gas to be $\sim$10
M$\sun$, and the SN explosion energy to be $\sim$10$^{50}$ erg.
From the morphology and spectral properties of G352.7$-$0.1, they
concluded that this remnant  belonged to the mixed-morphology (MM)
class. \citet{toledo-Roy2014} explored a blowout scenario to
explain the morphological features of G352.7$-$0.1, presented 3D
hydrodynamical simulations of SNR.

Recently, \citet{pannuti2014} presented X-ray study of the remnant
by using data from {\it XMM-Newton} and  {\it Chandra}. Their {\it
XMM-Newton} spectra gave the best-fit with a single thermal
component ($kT_{\rm e}$ $\sim$ 1.2 keV) in NEI condition with
enhanced abundances of Si and S. On the other hand, their {\it
Chandra} spectra has been best described with two thermal
components (VNEI+VNEI) with temperatures of $\sim$2.14 keV for
hard and $\sim$0.39 keV for soft component for the whole remnant,
although the spectra of some regions can be fit with single VNEI
model. They obtained overabundant Si and S and confirmed that the
X-ray emission comes from ejecta. They also presented the
detection of infrared emission from this SNR at 24 $\micron$ by
MIPS aboard {\it Spitzer}.

G352.7$-$0.1 is one of Fe-rich SNRs in our Galaxy. Although it has
been previously argued to be a core-collapse (CC) SN origin by
\citet{giacani2009} and \citet{pannuti2014}, its progenitor has
not been well identified yet. Since {\it Suzaku}
\citep{mitsuda2007} has better sensitivity to Fe K-shell lines
than other satellites \citep{yamaguchi2014}, our main aim is to
identify the origin of G352.7$-$0.1 by determining Fe abundance
based on a deep observation ($\sim$201.6 ks) with the X-ray
Imaging Spectrometer \citep[XIS;][]{koyama2007} onboard {\it
Suzaku}.

The rest of this paper is organized as follows: We present the
observation and the data reduction in Section 2. We explain the
image and spectral analysis in Section 3. In section 4, we discuss
the results and in Section 5, we summarize our conclusions. All
through this paper, the errors are given at 90\% confidence level.

\section{Observation and Data Reduction}
G352.7$-$0.1 was observed with the XIS on the focal plane of the
X-ray telescope \citep[XRT;][]{serlemitsos2007} onboard {\it
Suzaku}, starting on 2012 March 02 at 20:39 (UT) and ending on
2012 March 07 at 13:17 (UT) under the observation ID 506052010.
The pointing position was ($l$, $b$)=($352\fdg7$, $-0\fdg12$). The
XIS was operated in the normal clocking mode (8 s exposure per
frame), with the standard 5$\times$5 or 3$\times$3 editing mode
\citep{koyama2007}. The XIS consists of four sets of X-ray CCD
camera systems (XIS0, 1, 2, and 3). XIS1 has a back-illuminated
(BI) sensor, while XIS0, 2, and 3 have front-illuminated (FI)
sensors. XIS2 has not been functional due to an unexpected anomaly
since 2006 November\footnote{
http://www.astro.isas.jaxa.jp/suzaku/doc/suzakumemo/
suzakumemo-2007-08.pdf}. Three out of the four CCD chips were
available in this observation: XIS0, XIS1, and XIS3. A fraction of
the imaging area of the XIS0 has also been unusable since 2009
June\footnote{http://www.astro.isas.jaxa.jp/suzaku/doc/suzakumemo/
suzakumemo-2010-01.pdf}. Each CCD chip contains 1024$\times$1024
pixels (1 pixel=24 $\mu$m$\times$24 $\mu$m) for a 17.8$\times$17.8
arcmin$^{2}$ field of view (FOV). Two calibration sources of
$^{55}$Fe are installed to illuminate two corners of each CCD for
absolute gain tuning.

We used High Energy Astrophysics Software ({\sc HEASoft})
package\footnote{http://heasarc.nasa.gov/lheasoft/} version 6.11.1
for the data reduction, {\sc xselect} version 2.4 to extract
images and spectra, and the X-ray Spectral fitting package ({\sc
xspec}) version 12.7.0 \citep{arnaud1996} for the X-ray spectral
analysis. The {\it Suzaku}/XIS calibration database (CALDB:
20130305)\footnote{http://www.astro.isas.ac.jp/suzaku/caldb/} was
updated in 2013 March. We downloaded the archival data from the
Data Archives and Transmission System
(DARTS)\footnote{http://www.darts.isas.jaxa.jp/astro/suzaku/}. We
used the cleaned event file created by the {\it Suzaku} team. For
the spectral analysis, we generated the Redistribution Matrix File
(RMF) and Ancillary Response File (ARF) by using the {\sc
xisrmfgen} and {\sc xissimarfgen} in the {\sc HEASoft} package,
respectively \citep{ishisaki2007}.

\section{Analysis}
\subsection{Imaging analysis}
In Figure 1, we show the XIS mosaic image of G352.7$-$0.1 in the
full energy band of 0.3$-$10.0 keV. To produce the X-ray image,
the data of XIS0, XIS1, and XIS3 were added by using {\sc ximage}
package, and the resulting image was smoothed with a Gaussian of
$\sigma$=3 arcmin. The solid and dashed circles show the spectral
extraction regions (see Section 3.2). We excluded calibration
sources at the corners of the CCD chips. X-ray image was overlayed
with the VLA radio continuum image (in green contours) of
G352.7$-$0.1 obtained at 4.8 GHz (E. Giacani, private
communication).

\begin{figure}[t]
\centering
\includegraphics[width=0.5\textwidth]{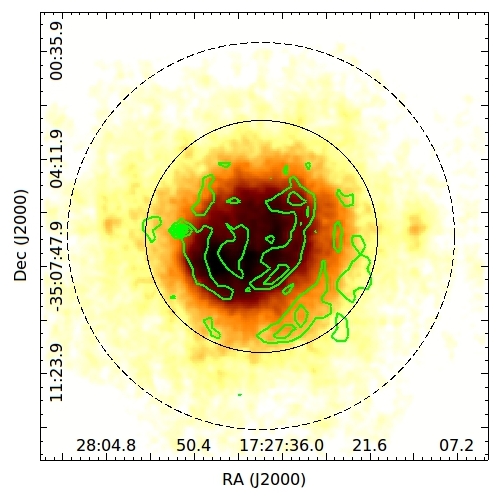}
\caption{\small{The X-ray image of G352.7$-$0.1 with the {\it
Suzaku} XIS in the full energy band 0.3$-$10.0 keV. The image was
smoothed with a Gaussian of $\sigma$=3 arcmin. The source and
background regions of the X-ray spectra are given by the solid
circle with a 3.9 arcmin radius and dashed annulus with a 3.9 and
6.5 arcmin radii, respectively. The radio image at 4.8 GHz band
obtained by the VLA observation \citep{giacani2009} is overlaid as
the contour image in green. The contour levels are 2.54, 5.48,
8.42, and 12.1 mJy beam$^{-1}$. The corners of the CCD chip
illuminated by $^{55}$Fe calibration sources are excluded from the
image.}}
\end{figure}

\subsection{Spectral analysis}
The XIS spectra of the source were extracted from a circular
region with a radius of 3.9 arcmin, centered at
$\rm{RA}(2000)=17^{\rm{h}} 27^{\rm{m}} 39^{\rm{s}}$,
$\rm{Dec.}~(2000)=-35\degr 06\arcmin 49\arcsec$ as shown in Figure
1. We extracted background spectrum taken from an annulus region
surrounding the remnant (the inner radius 3.9 arcmin and the outer
radius is 6.5 arcmin) in the same FOV away from the corners
containing calibration sources (see Figure 1). We note that the
angular size of the remnant is small enough to choose background
around it. The spectra were binned to a minimum of 30 counts per
bin to allow use of the $\chi^{2}$ statistic.

We first fitted the spectra with a NEI plasma model \citep[VNEI
model with neivers 2.0 in {\sc xspec};][]{borkowski2001} modified
by interstellar photo-electric absorption \citep[the wabs
component in {\sc xspec};][]{morrison1983}. The electron
temperature ($kT_{\rm e}$), the absorbing column density ($N_{\rm
H}$), the ionization time-scale ($\tau$=$n_{\rm e}t$), where
$n_{\rm e}$ and $t$ are electron density and the time since the
plasma was heated, and the normalization are free parameters,
while all elements were fixed at solar abundances
\citep{anders1989}. This model did not give an acceptable fit with
a $\chi^{2}$=2509.5 for 1018 degrees of freedom (d.o.f.). Then,
the elemental abundances of Si, S, Ar, Ca, and Fe were also let
free while other elemental abundances were fixed to their solar
values. In this case, the absorbed VNEI model provided a good fit
for the data, with a $\chi^{2}$ =1300.9 for 1013 d.o.f, but failed
to reproduce the Fe-K line profile. Residuals of the VNEI spectral
fit showed that there was clear residual emission at $\sim$6.4
keV. Therefore, we added a Gaussian component (gauss model in {\sc
xspec}) to the VNEI model and obtained a statistically acceptable
fit with a reduced $\chi^{2}$ of 1.1 for 1012 d.o.f. We note that
the Gaussian line width parameter was fixed to 0 eV. This fit gave
an ionization time-scale of $\sim$$2.3\times10^{10}$ cm$^{-3}$s,
indicating that the plasma is still ionizing. To understand the
origin of the Fe line (fluorescence or ejecta), we applied two
component NEI model wabs$\times$(VNEI+VNEI). In this fitting, the
electron temperature, ionization time-scale, normalization of both
components were let free. The abundances of all elements were
fixed at their solar values. This model did not give an acceptable
fit with a $\chi^{2}$=2509.4 for 1015 d.o.f. Then, the abundances
of Si, S, Ar, and Ca in first component, Fe in second component
were free parameters while the others were fixed to their solar
values. The fit was significantly improved with a $\chi^{2}$ of
1.02 for 1010 d.o.f and furthermore yielded the abundance of Fe.
Figure 2 shows FI (XIS0 and XIS3) spectra of the G352.7$-$0.1
fitted simultaneously through an absorbed VNEI+VNEI model in the
1.0$-$8.0 keV energy band. The best-fit parameters are given in
Table 1.

We also calculated the ratios of S, Ar, Ca, and Fe relative to Si
and compared them with the Type Ia SN models; carbon deflagration
\citep[W7;][]{nomoto1997}, delayed detonation
\citep[WDD2;][]{nomoto1997}, delayed detonation
\citep[DDTe;][]{badenes2003}, and pulsed delayed-detonation
\citep[PDDe;][]{badenes2003}. We also compared our results with CC
SN models \citep{woosley1995} with different progenitor masses of
11, 12, 15$M_{\sun}$ and presented them in Table 2.

\begin{figure}[t]
\centering
\includegraphics[width=0.5\textwidth]{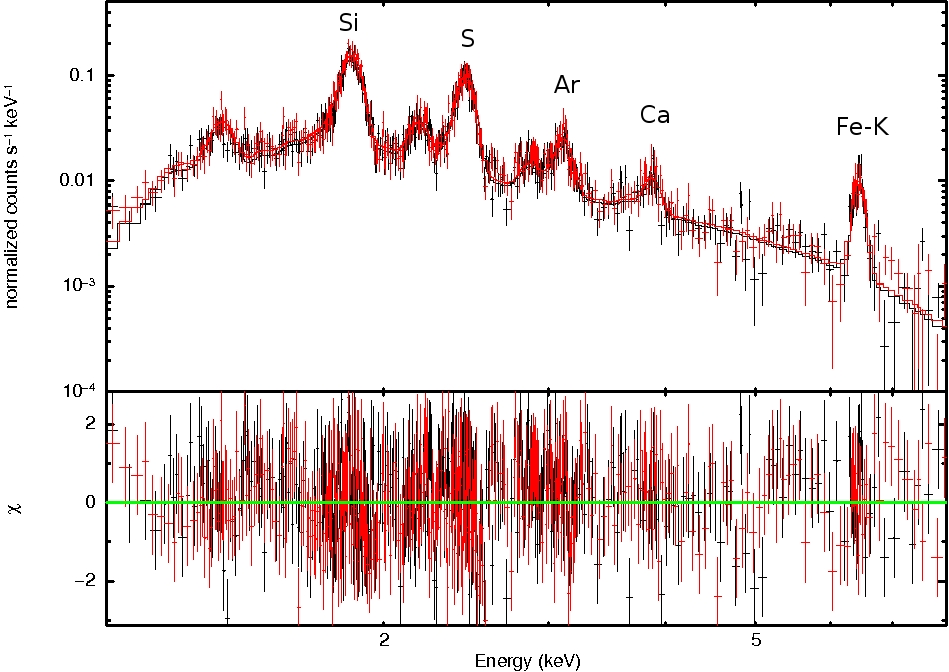}
\caption{\small{The FI spectra of G352.7$-$0.1 in the 1.0$-$8.0
keV energy band was fitted with an absorbed VNEI plus VNEI model.
The black and red data points represent XIS0 and XIS3 spectra,
respectively. Residuals from the model are plotted in the lower
panels.}}
\end{figure}

\section{Results and Discussion}
In this paper, we analyzed the X-ray imaging and spectroscopy of a
deep {\it Suzaku} observation of G352.7$-$0.1. The spectra were
well fitted with two-temperature thermal components; the softer
component has a plasma temperature of $\sim$0.6 keV, while the
harder component has plasma temperature of $\sim$4.3 keV in NEI.
The spectral analysis of this remnant gives us overabundance of Si
(4.3$\pm$0.4), S (7.0$\pm$0.8), Ar (11.1$\pm$1.6), Ca
(18.8$\pm$4.3), and Fe (6.8$\pm$0.7) and this state of
overabundance confirms that thermal X-ray emission is dominated by
ejecta.

The X-ray morphology of G352.7$-$0.1 was investigated by previous
studies with {\it ASCA}, {\it XMM-Newton}, and {\it Chandra}
observation. Using {\it ASCA} observation \citet{kinugasa1998}
proved that the X-ray emission of the remnant is a shell-type
morphology similar to that observed in the radio band. However,
{\it XMM-Newton} image of the SNR showed that the X-ray emission
completely fills the interior of the radio remnant; therefore,
\citet{giacani2009} concluded that this remnant belonged to the MM
class. \citet{pannuti2014} confirmed that this SNR is MM by using
{\it XMM-Newton} and {\it Chandra} data. MM SNRs \citep{rho1998},
also known as thermal composites, are identified with shell
emission in the radio band and centrally brightened thermal
emission in the X-ray band with little or no limb brightening. As
seen in Figure 1, XIS image of the G352.7$-$0.1 in the 0.3$-$10.0
keV energy band confirms that the SNR has a center-filled X-ray
plasma within the radio shell. Although there have been several
models to explain MM SNRs (e.g. evaporation model;
\citealp{white1991} and thermal conduction model;
\citealp{cox1999}), they are not well understood. Both of the
above models suggested that the X-ray emission of MM SNRs is
related to the shocked ambient medium, while \citet{slane2002}
were the first to propose that ejecta emission can be
non-negligible. The X-ray emission from the ejecta is significant
for several MM SNRs \citep[e.g.][]{lazendic2006}. Our analysis of
{\it Suzaku} XIS data confirmed that G352.7$-$0.1 has both MM and
ejecta structure. Generally, MM SNRs expand in very dense regions.
Most of them interact with molecular clouds and are detected in
the $\gamma$-ray wavelength. However, neither an OH (1720 MHz)
maser nor TeV/GeV $\gamma$-ray emission has been detected from
this remnant.

MM SNRs are generally in association with molecular clouds, which
supports a CC origin for them. However, some MM SNRs have Type Ia
origin, such as G272.2$-$3.2 \citep{sezer2012, sanchez-Ayaso2013,
mcEntaffer2013} and G344.7$-$0.1 \citep{yamaguchi2012}. Moreover,
G344.7$-$0.1 was previously suggested to be a CC SNR;
\citet{yamaguchi2012}, however, showed that G344.7$-$0.1 was
originated from a Type Ia explosion by using {\it Suzaku} data.
The progenitor of MM SNR G352.7$-$0.1 has not been well identified
yet. \citet{giacani2009} favor an origin of a Type II SN, based on
the interactions of the SNR with an asymmetric wind. Using {\it
Chandra} observation, \citet{pannuti2014} argued that it might be
a CC, not a Type Ia SN, considering the possibility of a compact
central object (CCO) associated with it. In a recent work,
\citet{yamaguchi2014} favored Type Ia origin for this remnant,
considering Fe K-shell at $\sim$6443 eV line obtained from {\it
Suzaku} data.

From {\it Suzaku} X-ray spectra of G352.7$-$0.1, we clearly
detected strong Fe K-shell emission which clearly discriminate the
progenitor type. One of typical characteristics of Type Ia SNRs is
that they contain Fe-rich ejecta in the low ionization state
\citep[e.g., RCW 86;][]{yamaguchi2011}, while the Fe ejecta in CC
SNRs is in highly ionized \citep[e.g., Cas A;][]{maeda2009} or
sometimes in overionized \citep[e.g., W49B;][]{ozawa2009} state.
The spectra of G352.7$-$0.1 were modelled by two thermal NEI
plasma components (VNEI 1 and VNEI 2) which are super-solar
abundant in heavy elements and clearly indicate the ejecta origin
of them. The soft component (VNEI 1) consists of Si, S, Ar, and Ca
while the hard component (VNEI 2) consists of only Fe element. The
Fe ejecta component is at a temperature $\sim$7 times higher and
an ionization age $\sim$40 times lower than the VNEI 1 component.
Considering strong Fe K-shell emission and its being in low
ionization state, we conclude that this remnant has been
originated from Type Ia explosion. The low ionization age of the
hard component indicates that the Fe ejecta was more recently
ionized by reverse shock than the Si, S, Ar, and Ca ejecta. In
some Type Ia SNRs, low ionization age of Fe ejecta has been
detected (\citealp[e.g., SN 1006:][]{yamaguchi2008a}; \citealp[RCW
86:][]{yamaguchi2008b}; \citealp[Tycho:][]{hwang1998}). This fact
is explained by the authors as follows: the Fe ejecta has been
heated by reverse shock more recently than the other elements,
since it concentrates toward the center of the SNR.

We compared the elemental abundance ratio of S/Si $\sim$ 1.6,
Ar/Si $\sim$ 2.6, Ca/Si $\sim$ 4.4, and Fe/Si $\sim$ 1.6 with the
Type Ia and CC models with different progenitor masses in
Subsection 3.2. As seen in Table 2, our abundance ratios are
roughly consistent with W7 model in S/Si and Fe/Si, but the
abundance ratios of Ca/Si and Ar/Si show a large inconsistency
from the model. We obtained a strong Ca abundance from the
G352.7$-$0.1 for the first time, which led to high Ca/Si ratio
similar to that found in SNR G337.2$-$0.7 \citep{rakowski2006}.
Our abundance ratio of Ca/Si is roughly close to PDDe and DDTe
models by \citet{badenes2003}, but Ar/Si is about 2-4 times higher
than Type Ia models, maybe because of the asymmetric distributions
of ejecta. Our abundance pattern seen in the {\it Suzaku} spectra
does not strongly match with Type Ia models.

We estimate some physical parameters of G352.7$-$0.1 based on the
best-fit emission measure, sum of two components
$\sim$$81.9\times$10$^{56}$ cm$^{-3}$. For this purpose, we assume
the distance to the SNR to be 7.5 kpc \citep{giacani2009} and the
plasma is a sphere with radius of $R$=3.9 arcmin. We calculate
X-ray emitting volume $V$ to be $\sim$7.6$\times$10$^{58}f$
cm$^{3}$ from $V$=$(4/3)\pi R^{3}f$, where $f$ is the filling
factor. Then, we estimate the electron density of the plasma
$n_{\rm e}$ to be $\sim$0.36$f^{-1/2}$ $\rm cm^{-3}$ (assuming
$n_{\rm e}$ $\sim$ 1.2n$_{\rm H}$). From $t$=$\tau$/$n_{\rm e}$ we
estimate the age of Fe ejecta plasma, which has a low ionization
time-scale of $\tau$ $\sim$ 8.8$\times$10$^{9}$ cm$^{-3}$ s, to be
$\sim$$780f^{1/2}$ yr, indicating that Fe ejecta has been heated
by reverse shock very recently. Finally, we calculate the mass of
the X-ray emitting plasma to be $M_{\rm x}$ $\sim$ $23f^{1/2}$
$M_{\sun}$ from equation $M_{\rm x}$=$n_{\rm e}m_{\rm H}V$, where
$m_{\rm H}$ is the mass of a hydrogen atom. The abundances were
consistent with super-solar values and derived low X-ray emitting
mass ($\sim$$23f^{1/2}$ $M_{\sun}$) indicate that the plasma is
dominated by ejecta material.

\section{Conclusion}
As a result of {\it Suzaku} XIS analysis, we found that the plasma
of G352.7$-$0.1 is thermal, heavily absorbed with a column density
of $\sim$3.3$\times$10$^{22}$ cm$^{-2}$, consists of a soft
component with a plasma temperature of $\sim$0.6 keV and a hard
component with a temperature of $\sim$4.3 keV in NEI condition.
The abundances of Si, S, Ar, Ca, and Fe are found to be enhanced
above the solar values confirming the ejecta-dominated nature of
G352.7$-$0.1. We conclude that the origin of the Fe-K emission is
Fe-rich ejecta at high-temperature plasma in a low ionization
state and the origin of SN explosion is Type Ia.

\acknowledgments We thank Elsa Giacani for providing us with the
4.8 GHz VLA data. AS is supported by T\"{U}B\.{I}TAK (The
Scientific and Technological Research Council of Turkey)
PostDoctoral Fellowship. FG acknowledges support by the Akdeniz
University Scientific Research Project Management. $~$

$~$

Facilities: \facility{Suzaku}.

\begin{table}
\caption{The Best-fit Spectral Parameters of G352.7$-$0.1 Through
the Use of {\it Suzaku} Data}
\begin{center}
 \begin{tabular}{@{}ccc@{}}
  \hline\hline
      Component&Parameters & Values \\
\hline
Absorption&$N_{\rm H}$ (10$^{22}$ cm$^{-2})$&3.3$\pm$0.1 \\
VNEI 1&$kT_{\rm e}$ (keV)&0.6$\pm$0.1 \\
&Si (solar)&4.3$\pm$0.4 \\
&S (solar)&7.0$\pm$0.8\\
&Ar (solar)&11.1$\pm$1.6 \\
&Ca (solar)&18.8$\pm$4.3\\
&$\tau$ (10$^{11}$ cm$^{-3}$ s)&3.4$\pm$0.2 \\
&$EM$\footnote{Emission measure $EM$=$\int n_{\rm e}n_{\rm H}$dV,
where $n_{\rm e}$ and $n_{\rm H}$ are the number densities of
electrons and protons, respectively and $V$ is the X-ray-emitting volume.}($10^{57}$ $\rm cm^{-3}$)&7.4$\pm$0.6\\
VNEI 2&$kT_{\rm e}$ (keV)&4.3$\pm$0.2 \\
&Fe (solar)&6.8$\pm$0.7\\
&$\tau$ (10$^{9}$ cm$^{-3}$ s)&8.8$\pm$0.9 \\
&$EM$($10^{56}$ $\rm cm^{-3}$)&7.9$\pm$0.4\\
 \hline
&$\chi^{2}/$d.o.f.$$&1034/1010\\
&reduced $\chi^{2}$&1.02\\
 \hline
{\bf Notes.} The errors are at
  90\% confidence level.
\end{tabular}
\end{center}
\end{table}

\begin{table}
\caption{Comparison of the Abundance Ratios with Type Ia Models
(\citealp[W7 and WDD2:][]{nomoto1997}; \citealp[PDDe and
DDTe:][]{badenes2003}) and CC Models \citep{woosley1995}}
\begin{center}
 \begin{tabular}{@{}ccccccccc@{}}
  \hline
    \hline
    &&\multicolumn{4}{c}{Type Ia Models}&\multicolumn{3}{c}{CC Models}\\
    \cline{3-7}
     \cline{7-9}
      Element Ratio& Best-fit Abundance Ratio&W7
      & WDD2& PDDe
      & DDTe&11$M_{\sun}$&12$M_{\sun}$&15$M_{\sun}$\\
\hline
S/Si& 1.63$\pm$0.21  &  1.07& 1.17 &1.5&1.4&0.87&1.53&0.62\\
Ar/Si& 2.58$\pm$0.24  &  0.89& 1.38 &0.68&0.60&0.63&1.62&0.50\\
Ca/Si& 4.37$\pm$0.32 &  0.75& 0.94 &2.9& 2.5&0.65&2.04&0.43\\
Fe/Si& 1.58$\pm$0.19  &  1.56& 0.85 &0.89&0.91&1.37&0.23&0.70\\
  \hline
\end{tabular}
\end{center}
\end{table}

\end{document}